\documentclass[iop,apj,preprint]{emulateapj}
\usepackage{amsmath,amssymb,amstext}
\usepackage{epstopdf}

\usepackage[breaklinks,colorlinks,citecolor=blue,linkcolor=magenta]{hyperref}

\usepackage{xcolor}

\usepackage[all]{hypcap} 

\shorttitle{The initial physical conditions of the Orion fingers}
\shortauthors{Rivera-Ortiz, P. R. et al.}

\begin{document}
\title{The initial physical conditions of the Orion BN/KL fingers}
\author{P. R. Rivera-Ortiz\altaffilmark{1,2},A. Rodr\'i guez-Gonz\'alez\altaffilmark{1,2}, L. Hern\'andez-Mart\'inez\altaffilmark{1}, J. Cant\'o\altaffilmark{3}, Luis A. Zapata\altaffilmark{4}}
\affiliation{$^1$ Instituto de Ciencias Nucleares, 
Universidad Nacional Aut\'onoma de M\'exico, 
Ap. 70-543, 04510 D.F., M\'exico}
\affiliation{$^2$LUTH, Observatoire de Paris, PSL, CNRS, UMPC, Univ Paris Diderot, F-92195 Meudon, France}
\affiliation{$^3$Instituto de Astronom\'{\i}a, Universidad
Nacional Aut\'onoma de M\'exico, Ap. 70-264, 04510 D.F., M\'exico} 
\affiliation{$^4$Instituto de Radiastronom\'ia y Astrof\'isica, UNAM, Apdo. Postal 3-72 (Xangari), 58089 Morelia, Michoac\'an, M\'exico}
\begin{abstract}
Orion BN/KL is an example of a poorly understood phenomena in star forming regions involving the close {encounter of young stellar objects}. The explosive structure, the great variety of molecules {observed}, the energy involved in the event and the mass of the region suggest a contribution in the {chemical diversity} of the {local} interstellar medium. Nevertheless, the frequency and duration of other events like this have not been determined. 
In this paper, we explore a recent analytic model that takes into account the interaction of a clump with its molecular environment.  We show that the widespread kinematic ages of the Orion fingers - 500 to 4000 years- is a consequence of the interaction of the explosion debris with the surrounding medium. This model explains satisfactorily the age discrepancy of the Orion fingers, and infers the initial conditions together with the lifetime of the explosion. Moreover, our model can explain why some 
CO streamers do not have a H$_2$ finger associated. 
\end{abstract}

%
\section{Introduction} 
\label{sec:intro}
Orion BN/KL is a complex massive star formation region that is associated with an explosive event that occurred some 500 years ago. In particular, it contains around 200 filamentary structures in H$_2$ emission known as the Orion fingers, which could be formed by the close {encounters of young stellar objects} \citep[][and references therein]{ZETAL09,BETAL11}. {The most accepted interpretation of these fingers is that they were formed by the interaction of high velocity gas clumps with the environment \citep{BETAL17}. We will consider this interpretation.}

The age of the event have been determined by several authors using different techniques. \cite{BETAL11} analyzed the projected position and velocity {of the heads} of the H$_2$ fingers. For each finger, they found an individual age that is between 1000 and 500~yr. This is in contradiction with the idea that Orion BN/KL was produced by a single explosive event and that the expelled clumps are in ballistic motion, so they concluded that there must be some deceleration.  \citet{ZETAL09} {reported the counterpart of the H$_2$ fingers observing the J$=2\to1$ CO transition, called CO streamers. Each streamer has a radial velocity that increases linearly with the distance to a common origin and, assuming a simultaneous ejection}, they determined the 3D structure and obtained a most probable age of approximately 500~yr.  This is in agreement with the age estimated by \cite{RETAL17}, who used the proper motions and projected positions of the runaway objects I, n and BN to estimate a close encounter 544 years ago. { Also, \cite{ZETAL11a} calculated the age of a expanding bubble in $^{13}$CO centered in the same possible origin of the region. The radial velocity and the size of this outflow result in $\sim600$ years}. The momentum and kinetic energy  of this outflow is at least 160~M$_\odot$ km s$^{-1}$ and $4\times 10^{46}$ and $4\times10^{47}$~erg \citep{SETAL84,KS76}. 

 There is a chance that the fingers could be originated at different moments. Perhaps, there is an unexplored mechanism to produce such an extended structure. The machine-gun model has been mentioned as a possible explanation, but previous models \citep{RB93}, even when they are not colimated, are far from being as isotropic as the Orion fingers. Then, the runaway stars \citep{RETAL17}, the expansion of the molecular bubble \cite{ZETAL11b} and the age determined by the CO  streamers \citep{ZETAL09}, are strong evidence of a single and simultaneous event. Then, the widespread ages could be explained by a dynamical model that takes into account the deceleration of a dense clump by the surrounding environment.

\begin{figure}[!ht]
     \centering
     \includegraphics[width=\columnwidth]{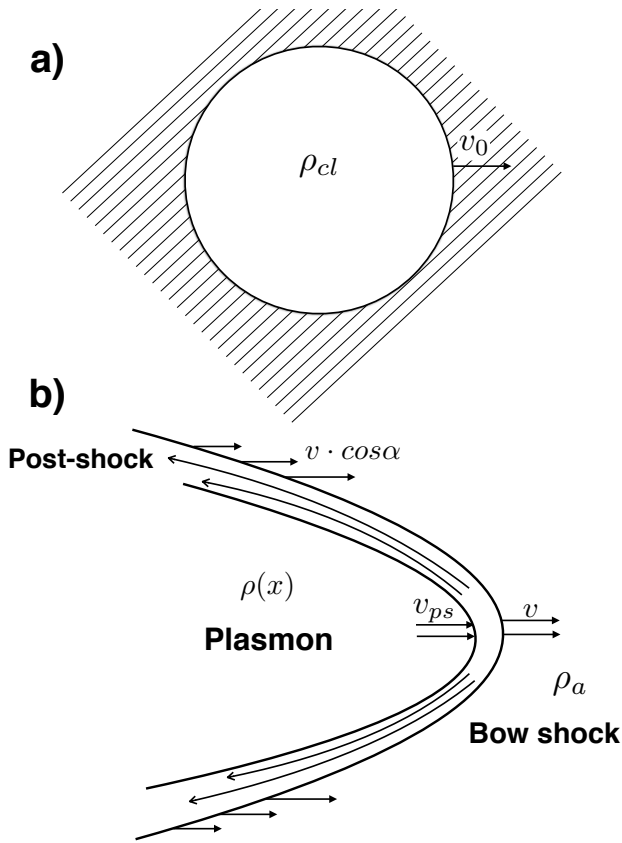}
     \caption{ {\bf a) Schematic representation of the initial clump at the ejection moment. The ejected clump takes a plasmon profile by the pressure balance between the internal pressure and the ram pressure produced by the velocity component $v\cos{\alpha}$, where $\alpha$ is the angle between the plasmon surface normal  and the motion direction. b) In our model (see RO19) the reverse shock deforms the initial clump that becomes into a plasmon in a negligible time. The environment has a density $\rho_a$, the plasmon has a velocity $v$ and a density $\rho (x)$ with a density structure studied in DA. The post-shock region that separates the environment and the plasmon structure has been exaggerated for clarity. An intermediate phase between this two cases was well studied by \cite{B97}  and \cite{BETAL15}.}}
     \label{fig:plasmon}
\end{figure}

 %
There are several attempts to describe the interaction of a moving cloud against a static medium. \cite{DYA67} (hereafter DA) analyzed the plasmon problem, which consists in a moving cloud that adopts a particular density structure, and derived its equation of motion. \cite{CETAL98} improved the plasmon solution including centrifugal pressure. Also, \cite{RETAL98} proposed the equation of motion of a static spherical cloud that is accelerated with a high velocity wind due to the ram pressure.
 More recently, \cite{ROETAL19} (hereafter RO19) proposed a modification to the plasmon problem, considering the mass lost by the clump, which can modify a plasmon dynamic history if it is embedded in a high density environment.  The plasmon problem is based on the direct consideration of the balance between the ram pressure of the environment and the internal, stratified pressure of the decelerating clump. Fig. \ref{fig:plasmon}  represents the plasmon profile adopted by the pressure balance, the post-shock region, where the material is ionized, and the inner neutral region. A similar representation has been proposed by \cite{B97}.
 
 Then the dynamical analysis of the motion of the Orion fingers could lead to a better understanding of the conditions that formed such a structure.  \cite{BETAL15} performed numerical simulations of the fingers using observational restraints and obtained a notable resemblance to the  actual fingers. Nevertheless, as they described, the interpretation of such simulations is limited since they used an adiabatic system, while, in reality, the cooling length is much shorter than the total length of the longest fingers. Therefore, more detailed numerical solutions and an adequate analytic model can be helpful to determine the physical conditions and, perhaps, the ejection mechanism of the fingers, which can be helpful to understand the relevance and duration of similar events in the star forming processes.

 Then, adopting an age of $t=544$~yr \citep{RETAL17}, we propose a model to obtain the physical conditions of the ejection.
The mass-loss plasmon has a implicit dependence on its own size and it can be used to find better restrictions on the ejection mechanism. \\
In Section~\ref{sec:parameters} we describe the sample of objects to be analyzed, in Section~\ref{sec:analytic} we present the estimation of the properties for the clumps before the explosive event that generated the Orion fingers in Orion BN/KL. We summarize our conclusions in Section~\ref{sec:conclusions}.
\section{Obtaining the physical parameters of the fingers}
\label{sec:parameters}

\subsection{Proper motions}
\label{sec:observations}
From \cite{LB00}, \cite{DETAL02} and \cite{BETAL11} we have obtained the proper motion of several features and  the projected positions for the reported data. In the follow paragraphs we describe with more detail how this was done. 

\begin{itemize}
     
\item \cite{LB00} analyzed the proper motions of 27 bullets, with emission in [Fe {\small II}], and 11 H$_2$ knots, using a time baseline of 4.2~yr (see Figure~\ref{fig:rVdatos}). 
From these 38 objects only 19 have proper motion vectors aligned with the position vectors with respect to IRc2, the possible origin of the explosive event. They used a distance to the Orion Nebula of $d=450$~pc \citep{GS89}, that is larger than the actually accepted $d=414$~pc \citep{METAL07}) which leads to overestimate the projected distance and proper motion of the data. We have corrected this effect for this paper. In general, they conclude that the farther features have larger proper motions, which is consistent with, at least, some kind of impulse with an age shorter than 1000 yr. However, it is interesting to note that they reported some H$_2$ knots as almost stationary,  but these are not included in the final analysis.

\item \cite{DETAL02} measured the proper motions of several HH objects in the Orion nebula. For the Orion BN/KL region they found 21 HH objects moving away from IRc2. As \cite{LB00}, they found that the larger objects are faster.  HH 210 is also a prominent feature that has a proper motion of almost 400~km~s$^{-1}$. The uncertainties lead them to fit an age of $1010\pm140$~yr. Even in this case, several objects are not in the range of 870 to 1150~yr. Also,they used a distance of 450~pc, that has been corrected in this work to 414~pc.

\item \cite{BETAL11} (see also, \citep{CPHD06}) obtained the proper motions of 173 fingers in H$_2$, but in this case there is no clear evidence for a linear dependence of the velocity on the projected distance. They only mentioned that the age of the event could be between 500 and 1000~yr, whether the simultaneous ejection assumption is maintained. { The three data sets are represented in Figure~\ref{fig:rVdatos}}.

\item Also, \cite{ZETAL09} analyzed the CO streamers that seem to be related to the fingers. These streamers are $\sim$2 times shorter and narrower than the fingers and each one follow a Hubble law. The kinematic age of each one could be related to the projection angle with respect to the plane of the sky, and assuming that the  explosion was isotropic they found that the most probable age is around 500~yr. \cite{BETAL17}, using ALMA, found more streamers and confirmed that these streamers has isotropic extension. This means that some of the CO streamers do not have associated fingers.

\end{itemize}

\subsection{Mass, density and size}

On the other hand, from \cite{RETAL17}, \cite{CPHD06}  and \citet{BETAL17} we have obtained the mass, density and size of several features and  the projected positions for the reported data. In the follow paragraphs we also describe with more detail how this was done.

\begin{itemize}
    
\item  Recently, \cite{RETAL17} has measured, with high precision, the proper motions of the objects I, BN and n. They found that these objects had to be ejected from a common origin $544\pm6$~yr ago.  This uncertainty does not take into account systematic effects, which can increase it up to $\pm25$~yr. In any case, 544 years is consistent with the age determined by the CO streamers of about 550 years. In this work, we assume this event to be the origin of the ejection of the material that created the fingers and the streamers.

\item  \cite{CPHD06} measured 8$M_\odot$ as the mass of the moving gas. We can use this estimate to find the upper limits for either the mass of an individual clump, or its size. Nevertheless, due to the complexity of the region there is an uncertainty of a factor two in this mass estimate.

\item For the mass, we assume that the observed moving gas corresponds, exclusively, to that of the ejected clumps. Since there are 200 fingers, then the average mass of each clump is simply $8/200=0.04 M_\odot$. An inferior limit for the clump mass is that calculated by \cite{AB93} and \cite{BA94} of $10^{-5}M_\odot$ based on the [Fe II] 1.64$\mu$m line flux and size.

\begin{figure}[!ht]
\centering
\includegraphics[width=\columnwidth]{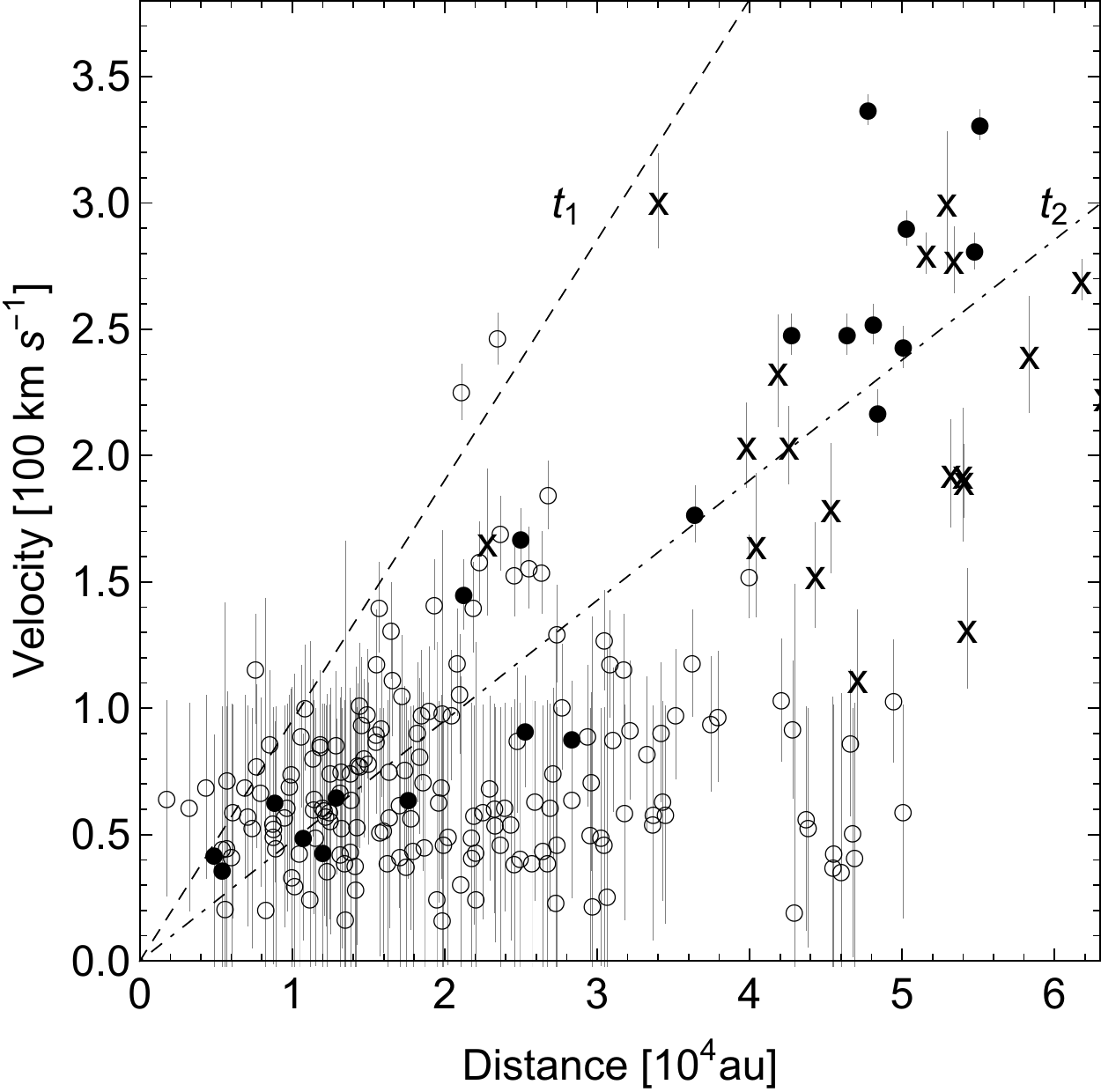}
\caption[Data sets used in this work to apply the dynamical models]{This figure shows the three data sets used for this work, with their respective uncertainties. The open circles stand for the H$_2$ fingers reported by \citet{BETAL11} see also, \cite{CPHD06}, the filled circles stand for the {[Fe\small II}] bullets \citep{LB00} and the crosses represent the HH objects reported by \cite{DETAL02}. The lines indicate an age consistent with no deceleration, $t_1=500$~yr (dashed) and $t_2=1000$~yr (dot-dashed). } 
\label{fig:rVdatos}
\end{figure}

\item On the other hand, an upper limit for the size of the initial clump is obtained by adopting the opposite assumption than above, that is, that all the moving mass comes from the swept up environmental material, and, a negligible amount from the clumps themselves. To follow this idea we have to fix the density of the environment. Extinction observations of the region by \cite{OHETAL16} and \citet{BETAL17} indicate densities between $10^5$ and $10^7$cm$^{-3}$. We adopt this latter limit, $n_a=10^7$cm$^{-3}$.
In reality, the density is highly structured (\cite{KETAL18}, \cite{BETAL87}). A better approximation would be to assume cylindrical symmetry for the Integral Spine Filament with a steep density gradient orthogonal to the spine. In this paper we assume an homogeneous environment, a cylindrical density profile would require to improve the presented plasmon dynamics.

\end{itemize}

\section{Analytic Model}
\label{sec:analytic}

We now model a finger as a cylinder of radius $R_{cl}$ and individual length $l_i$. Thus, the mass swept up by all the fingers (assuming the same radius) is,


\begin{equation}
    M_t=\pi R_{\rm cl}^2\mu m_h n_a \sum_i l_i,
\end{equation}
where $\mu=2$ is the mean molecular mass, $m_h$ mass of hydrogen and $n_a$ is the numerical density of the ambient medium. Considering, as a limit, that $M_t=8M_\odot$ is equal to the accelerated mass we can obtain  R$_{\rm cl}\sim90$~au, then this is the upper limit for the initial size of the ejected clumps.

\subsection{Ballistic motion}
The simplest model is to suppose that every ejected clump travels with constant velocity and, therefore, the motion is described by:\\
\begin{equation}
    r=vt.
    \label{eq:vcons}
\end{equation}

Since the projected length, $r$, and the velocity, $v$, also in projection, are observational data, then, the age of each clump can be obtained straightforward:
\begin{equation}
    t=\frac{r}{v},
    \label{eq:age}
\end{equation}
which is independent of projection.
\begin{figure}[!ht]
\centering
\includegraphics[width=\columnwidth]{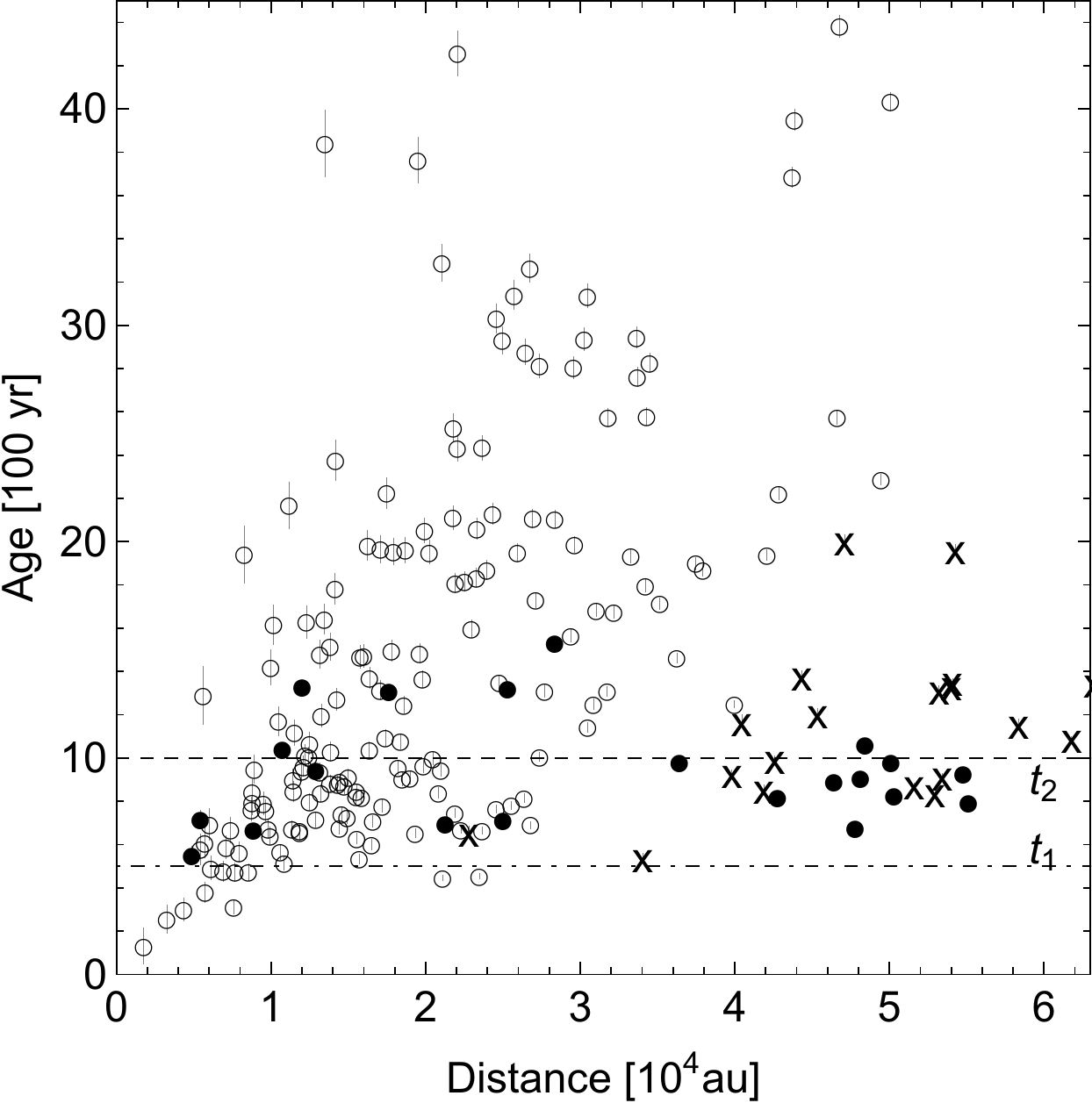}
\caption[Kinematic age assuming no deceleration]{Kinematic age assuming no deceleration. The symbol notation is the same as in Figure~\ref{fig:rVdatos}. The dashed line correspond to an age of 1000~yr and the dot-dashed line represents an age of 500~yr.} 
\label{fig:rT0}
\end{figure}

Therefore each clump has an individual age and { if} we assume that all of them were ejected in a single event, each age should be, at least, similar. { This is far from which we observe}. In Figure~\ref{fig:rT0} we show the result of  Equation~\ref{eq:age} applied to each data. The calculation of the spread of the error for the age was done using the standard procedure. The reported errors for the velocities of all the HH objects is 
10~km~s$^{-1}$ \citep{DETAL02}, of all the H$_2$ fingers is the 25~km~s$^{-1}$ \citep{CPHD06} and for the [FeII] bullets is reported in \cite{LB00} for each of them. Then, Figure~\ref{fig:rT0} implies that there was no simultaneous event or that the ballistic motion model is not an appropriate assumption. Deceleration is the most likely interpretation.

 Notice that the plasmon model assume an early interaction of the original clump with the environment that will modify its initial characteristics quickly (shape, density stratification or sound speed) to those of a plasmon. But the ram pressure prevents the plasmon's free expansion, and this effect gives shape to the material (see also, \citep{ROETAL19}, Figure 1).
\subsection{Dynamic model}
In order to determine the fundamental parameters that control the dynamics of a high velocity clump, such as the ejection velocity $v_0$, the { initial} size of the clump R$_{\rm cl}$, the density of the ejected material $\rho_{cl}$ and the density of the environment $\rho_a$, or their initial density contrast $\beta=\sqrt{\rho_a/\rho_{cl}}$, we use an analysis based on the plasmon proposed by DA. Assuming a spherical clump at the ejection, the initial mass can be expressed as, 
\begin{equation}
 M_0=\frac{4\pi R_{\rm cl}^3 \rho_{cl}}{3}=\frac{4\pi R_{\rm cl}^3 \rho_a}{3\beta^2}.
 \label{eq:mass}
\end{equation}
We assume that every clump was ejected with the same size ($R_{cl}=90$au) and the environment density is $10^7$~cm$^{-3}$, 
therefore we can estimate the ejection conditions. The plasmon density is not constant because of the enlargement of the traveled distance and the mass detachment included in the model. 

In this section we explore a model which takes into account the deceleration of the clump as it losses mass due to the interaction with the environment. This is the model developed in RO19. As stated in RO19, no matter the physical characteristics of the original clump (shape, size, density, velocity or temperature) the initial interaction of the clump with the surroundings will transform it into a plasmon as proposed by DA, \citet{CETAL98} and RO19. Mass, on the other hand, is preserved.

RO19 shows that the mass $M$, velocity $v$, and position $R$ of the newly created plasmon after a time $t$ of ejection/formation are given by the parametric form

\begin{equation}
{M}=M_0 e^{-\alpha \left(1-\frac{v}{v_0} \right)},
\label{eqn:lnm2}
\end{equation}
\begin{equation}
t={t_0}\int_{v/v_0}^{1} {u}^{-2/3}e^{-\frac{\alpha}{3} \left(1-{u} \right)} du,
\label{eq:fulltau}
\end{equation}
and
\begin{equation}
R=v_0t_0\int_{v/v_0}^1 u^{1/3}e^{-\frac{\alpha}{3}\left(1-u \right)}du,
\label{eq:r}
\end{equation}

respectively, where $M_0$ is the initial mass of the clump, $v_0$ the ejection velocity, {\bf $u=v/v_0$ is a dimensionless velocity}, $\alpha$ a parameter given by, 

\begin{equation}
\label{eq:lambda}
\alpha=\frac{8\lambda}{\pi+2}\sqrt{\frac{2}{\gamma-1}}\left(\frac{1}{\beta}\right),
\end{equation}

and a scale time $t_0$ 
\begin{equation}
    t_0=\frac{R_{\rm cl}}{\beta^2}\left(\frac{16\pi}{3\xi_{DA}(\gamma-1)^2} \right)^{1/3}\frac{1}{v_0},
    \label{eq:t0}
\end{equation}

with $\xi_{DA}=9.22$ from the DA model, $\lambda=0.0615$, and $\gamma=1.4$ is the adiabatic coefficient for an ideal diatomic gas.

\begin{figure}[!ht]
\centering
\includegraphics[width=\columnwidth]{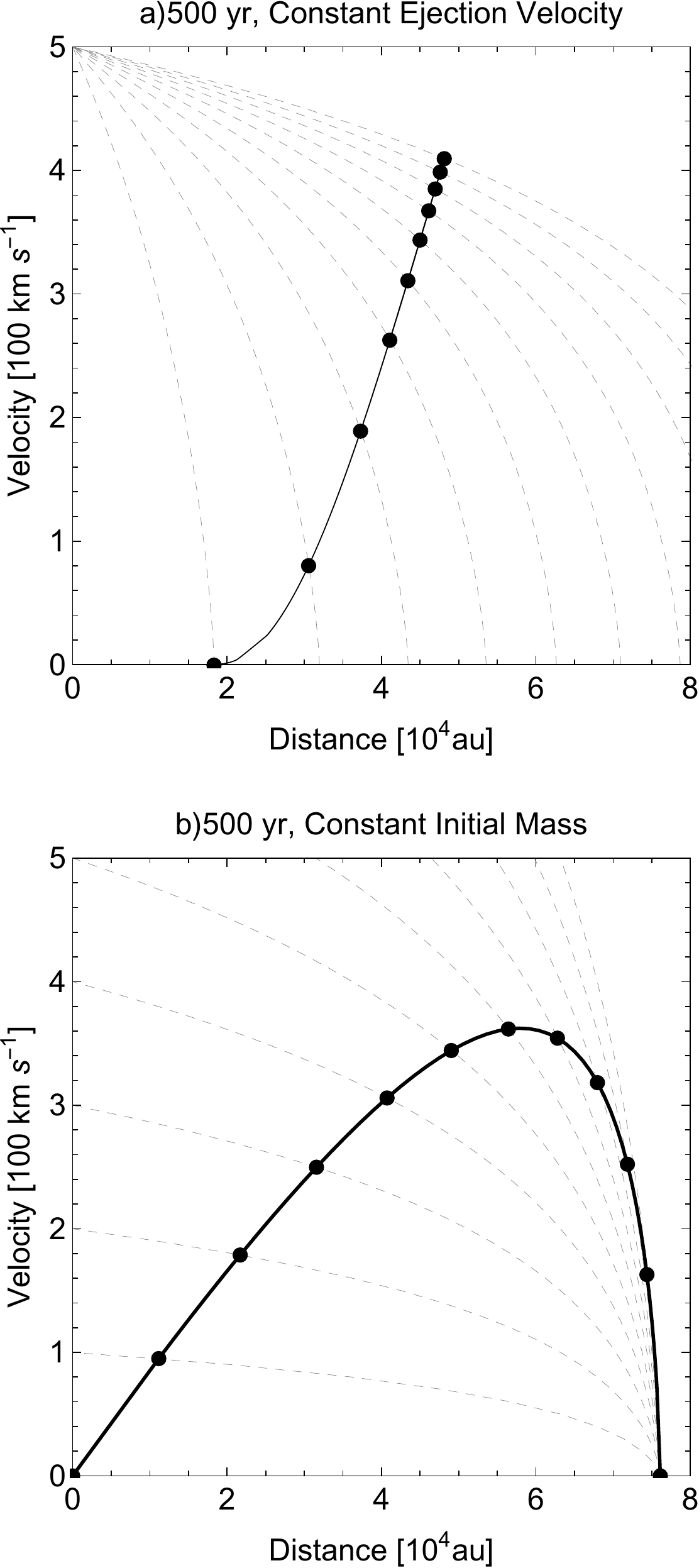}
\caption[Trajectories for a constant ejection velocity or constant ejection mass at a fixed time]{In both panels, the gray dashed lines are the trajectories for RO19 plasmons with a) different mass and constant ejection velocity 
(with a lower and higher mass trajectories of 2$\times$10$^{-2}$~M$_{\odot}$ and 2$\times$10$^{-1}$~M$_{\odot}$, respectively, 
divided into 10 equal intervals) and b) different ejection velocities and initial clump mass fixed at $M_0=0.2M_\odot$ (with a lower and higher velocities 
trajectories from 100~km~s$^{-1}$ to 1100~km~s$^{-1}$ with intervals of 100~km~s$^{-1}$). Given a fixed time $t=500$~yr, each 
trajectory reaches a position $R$ and a velocity $v$, marked as a black point in it.} 
\label{fig:curves}
\end{figure}

 Combining Equations (\ref{eq:lambda}) and (\ref{eq:t0}), we obtain:
 \begin{equation}
     \left[\frac{v_0}{\rm{km~ s}^{-1}}\right] \left[\frac{t_0}{\rm{yr}}\right]=233 \left[\frac{R_{\rm cl}}{\rm{au}}\right] \alpha^2.
     \label{eq:v0t0alfa}
 \end{equation}
 The purpose of the present paper is to use Equations (\ref{eq:mass}) to (\ref{eq:v0t0alfa}) to estimate the physical parameters, such as mass, ejection velocity, density, of each of the original clumps that produce the fingers we see today and formed by the interaction of the clumps with the surrounding molecular cloud.

We begin by assuming that all the clumps were ejected in a single explosive event that took place 544 years ago from the place of the closest interaction that expelled BN, n and I objects reported by \cite{RETAL17}. So, in Equation (\ref{eq:fulltau}) we set $t=544$yr for all the clumps, although each clump had their own initial mass and ejection velocity.

Next, for each clump we know, from observations, its distance to the origin of the explosion $R$ and its current velocity $v$. Both quantities are those on the plane of the sky. However, we take them as estimates of the real values, since there is no way to de-project them without making further assumptions.

Even so, we need to make a further assumption, since we have more unknowns than equations. We might, for instance, choose to assume a fixed value of $\beta$, which means the same initial density for each clump, or, perhaps, the same initial mass, or any other reasonable constrain. We choose, however, to assume a unique initial radius for all the clumps of $R_{cl}=90$au, based on the assumption that all the clumps were produced by the close encounter of two protostellar objects that ripped off material with the same cross section interaction. 

Then, we have a set of equations 
(equations~\ref{eqn:lnm2}, \ref{eq:fulltau} and \ref{eq:v0t0alfa}) that can be solved for $v_0$, $t_0$ and $\alpha$ simultaneously, 
and by Equation~(\ref{eq:mass}) we also can obtain the mass of each ejected clump. The number density of the surroundings was taken $n_a=10^7$cm$^{-3}$. {In Figure \ref{fig:curves} we show the trajectories of clumps in the $v-R$ plane as calculated by our model, using Eq. (5) to Eq. (10). A fixed clump radius $R_{cl}=90$au was assumed in all the calculations. In the upper panel, we have taken a fixed initial velocity for a clump with $v_0=500$km~s$^{-1}$, and vary its initial mass from $2\times10^{-2}$ (the lower dashed line) to $2\times10^{-1}M_\odot$ (the upper dashed line). The solid line marks the time $t=500$yr after ejection. In the bottom panel, the initial clump mass is also fixed at $M_0=0.2M_\odot$ and each dashed line corresponds to a different initial velocity $v_0$, from $100$ to $1100$km~s$^{-1}$. The solid line, again, marks the time $t=500$ yr after ejection. Note that clumps stop at the same distance, in this case at $75000$au.}

\begin{figure}[!ht]
\centering
\includegraphics[width=\columnwidth]{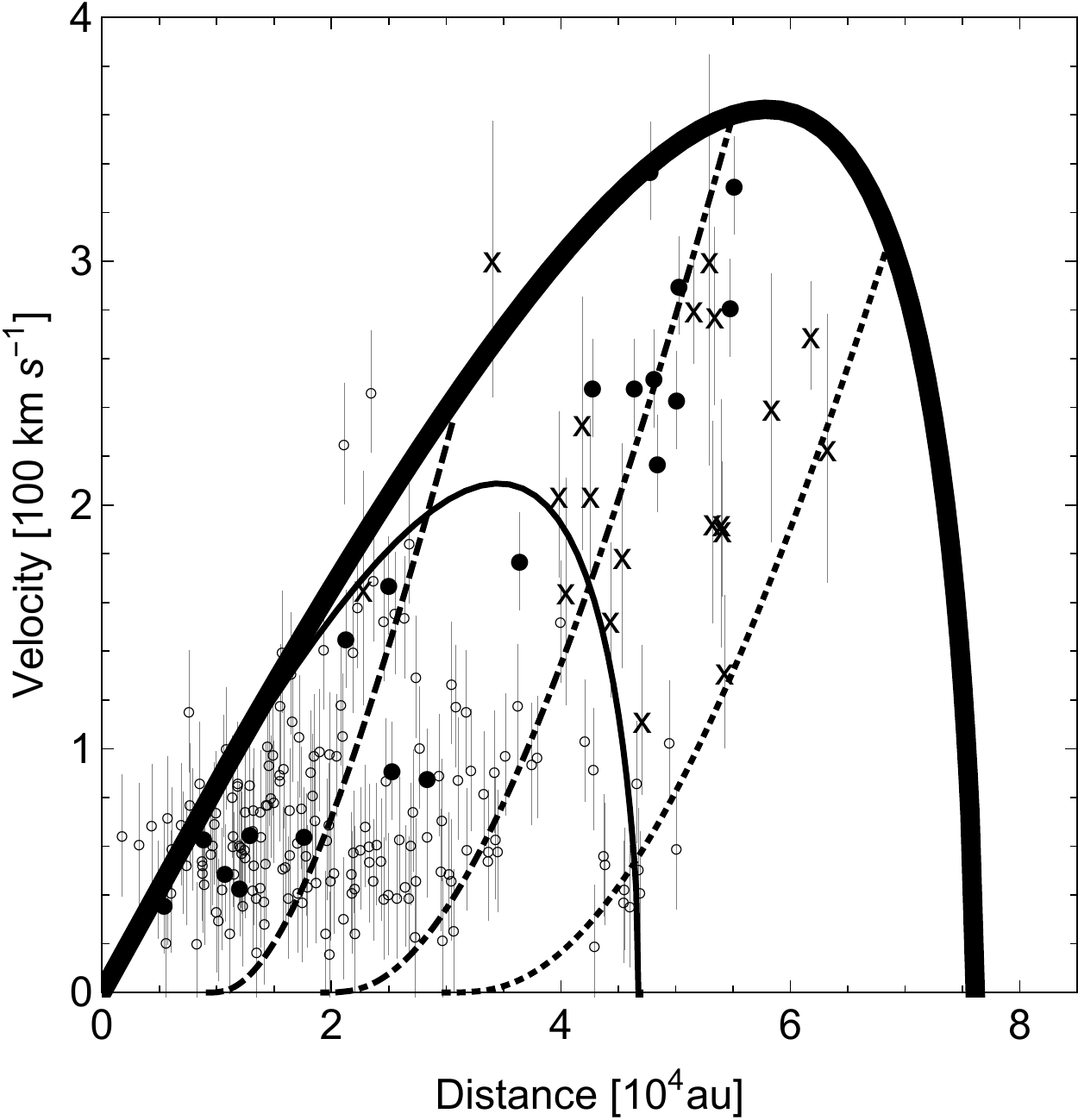}
\caption[RO19 model applied to the data sets]{In the figure are the data sets described in Figure~\ref{fig:rVdatos}, along with the fixed time curves with constant 
ejection velocities of $v=200, 500 $ and $ 800$~km~s$^{-1}$  (dashed, dot-dashed and dotted lines, respectively) and $M_0=0.2$ and $0.1$~M$_\odot$ constant mass (black thick and thin lines, respectively) using the RO19 plasmon model} 
\label{fig:ROmodel}
\end{figure}

In Figure~\ref{fig:ROmodel}, we can see that the model curves that envelope the data set do not have high mass ($>0.2$~M$_\odot$) and high velocity clumps ($>800$~km~s$^{-1}$). We could expect slow points with low velocities at a distance greater than $8\times10^4$~au, but there is not any evidence of such clumps but in this case we have that 800~km~s$^{-1}$ is the fastest velocity that meets the longer features. 
Also, a plasmon with ejected mass of $0.2$ M$_\odot$ will reach a final distance of $\sim8\times10^4$~au. This 
means that a less massive plasmon, with less than 800~km~s$^{-1}$ could be near to its lifetime or maybe 
it has already stopped. This could explain the CO streamers that are not related to any H$_2$ finger.

\begin{figure}[!ht]
\centering
\includegraphics[width=\columnwidth]{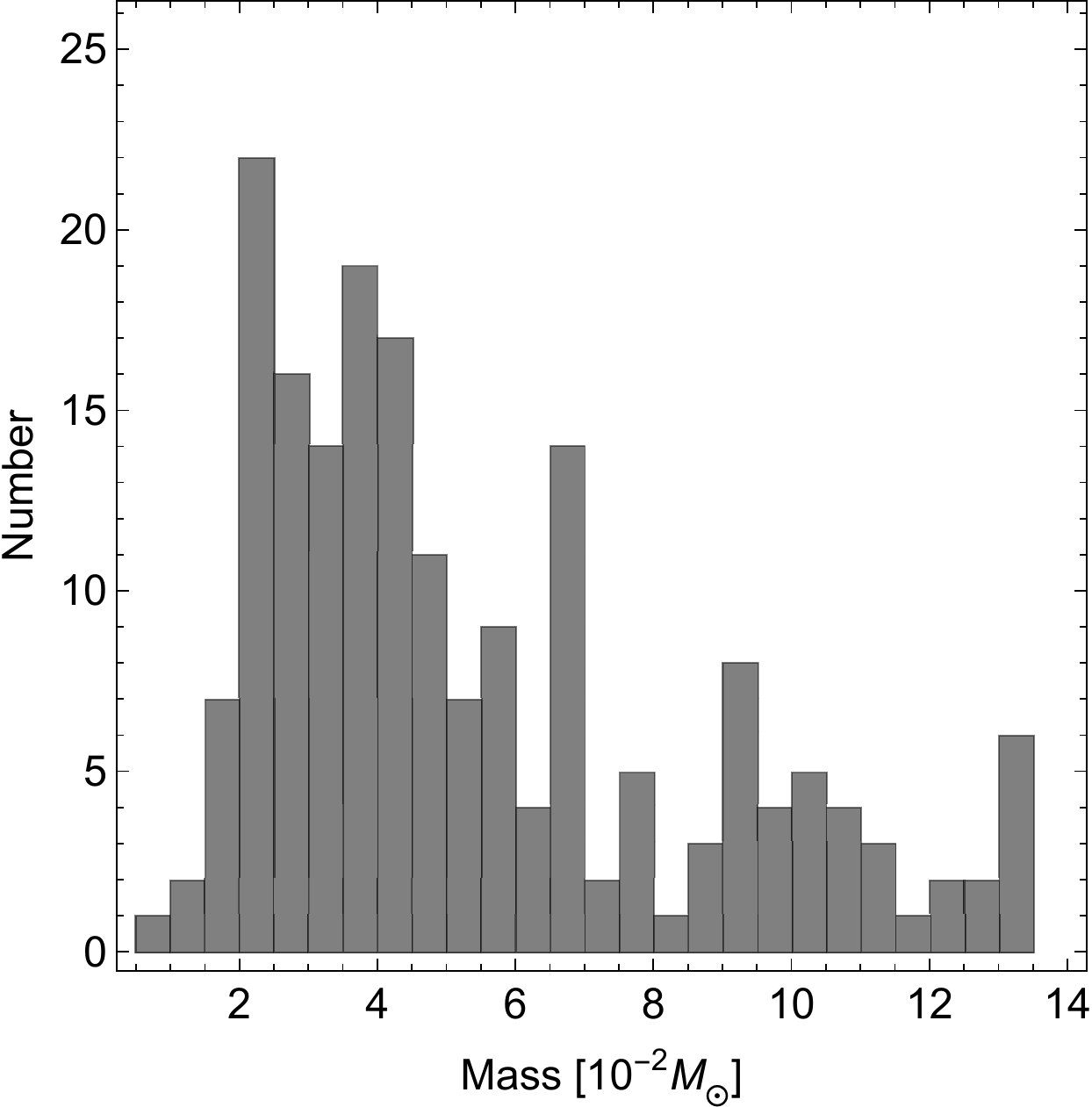}
\caption[Mass distribution]{The mass of the clumps calculated using the loosing mass plasmon model of RO19, for the data set presented in Section~\ref{sec:observations}} 
\label{fig:histomass}
\end{figure}

\begin{figure}[!ht]
\centering
\includegraphics[width=\columnwidth]{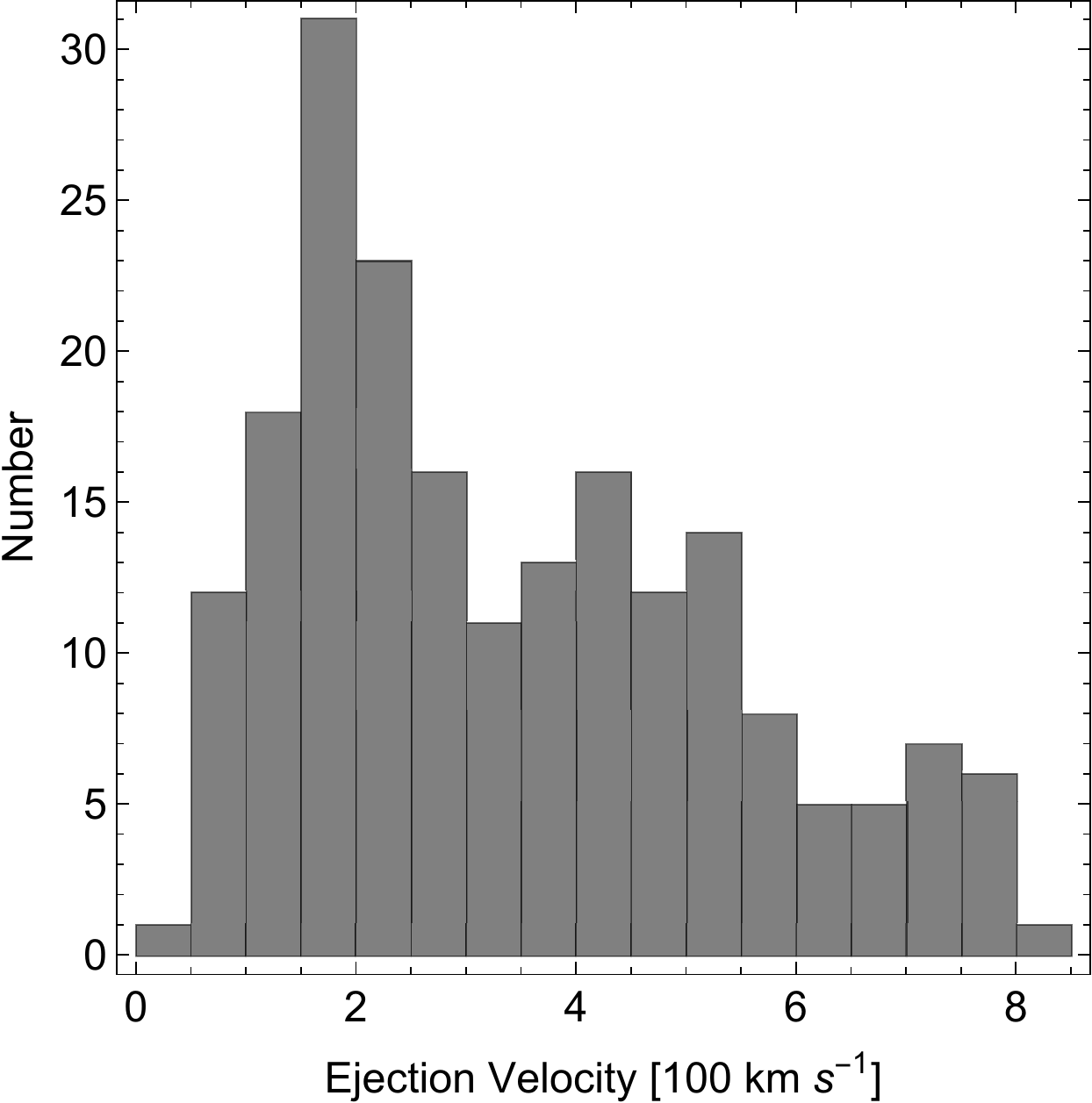}
\caption[Velocity distribution]{Velocity distribution according to the loosing mass plasmon model (see RO19), using the corresponding calculated ejection conditions.} 
\label{fig:histovel}
\end{figure}

Finally, the RO19 plasmon solution is applied to each of the object 
of the data sets of the  Sect.~\ref{sec:observations}  and the initial mass, ejection velocity and lifetime are obtained and shown in 
Figure~\ref{fig:histomass}, \ref{fig:histovel} and \ref{fig:histolife}, respectively. The total mass, Figure~\ref{fig:histomass}, is $11.93$~M$_\odot$ with mean mass of $0.06$~M$_\odot$ which is close to the limits of $4\times10^{-2}$~M$_\odot$ analyzed in Section \ref{sec:observations}.

Figure~\ref{fig:histovel} shows the ejection velocity distribution. It is interesting to note that there are 2 peaks in this distribution around 200 and 500~km~s$^{-1}$. Further analysis is required to propose a mechanism of explosion that could explain this characteristic. Also, the total kinetic energy of the model is $3\times10^{49}$~erg. 

Once the ejection parameters are obtained, we can infer the lifetime and stopping distance of each clump using $v=0$ in Equations~(\ref{eq:fulltau}) and (\ref{eq:r}). In Figure~\ref{fig:histolife} 
we show the distribution of the lifetime for the clumps. This can give an idea of the lifetime of the explosive event, in this case 2000~yr after the explosion, there will be just a few fingers and this can the reason why there are just a few cases of encounters of this kind.

\begin{figure}[!ht]
\centering
\includegraphics[width=\columnwidth]{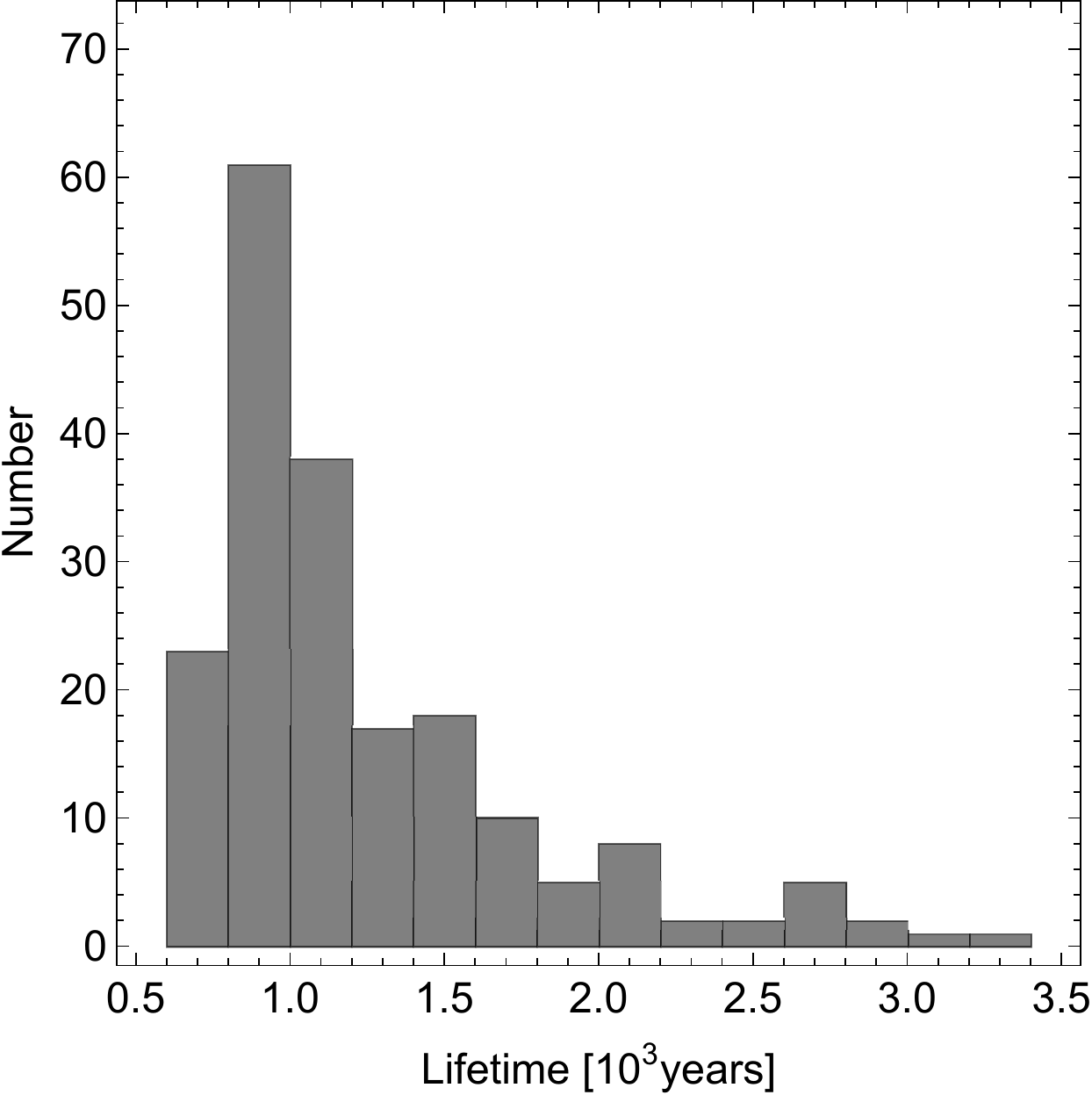}
\caption[Lifetime of the fingers]{Lifetime of each finger (for the sample used in this chapter) using R$_{\rm cl}=90$~au.} 
\label{fig:histolife}
\end{figure}

\begin{figure}[!ht]
\centering
\includegraphics[width=\columnwidth]{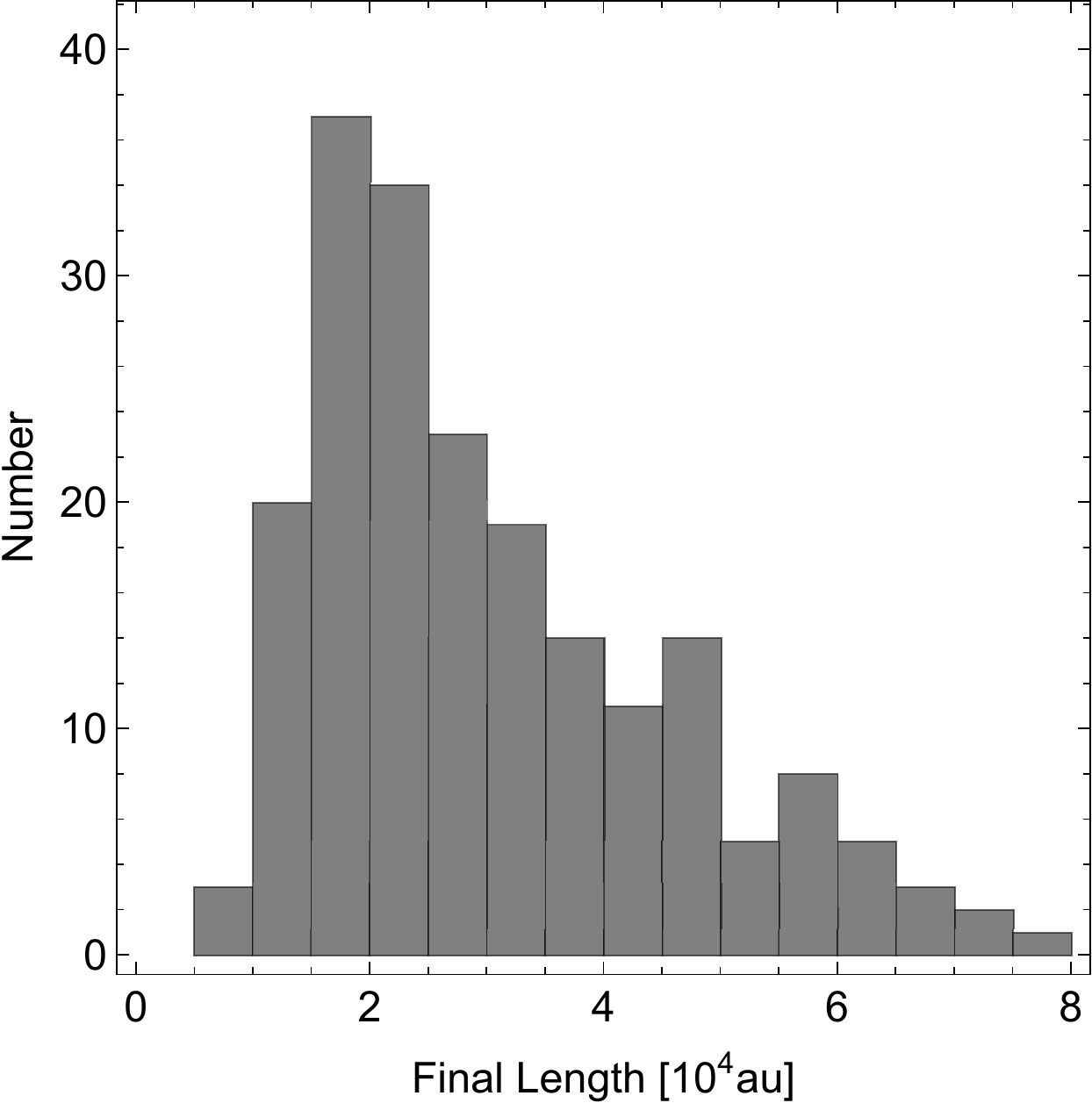}
\caption[Lifetime of the fingers]{Final length of each finger (for the sample used in this chapter) using R$_{\rm cl}=90$~au.} 
\label{fig:histolife}
\end{figure}

\begin{figure}[!ht]
\centering
\includegraphics[width=\columnwidth]{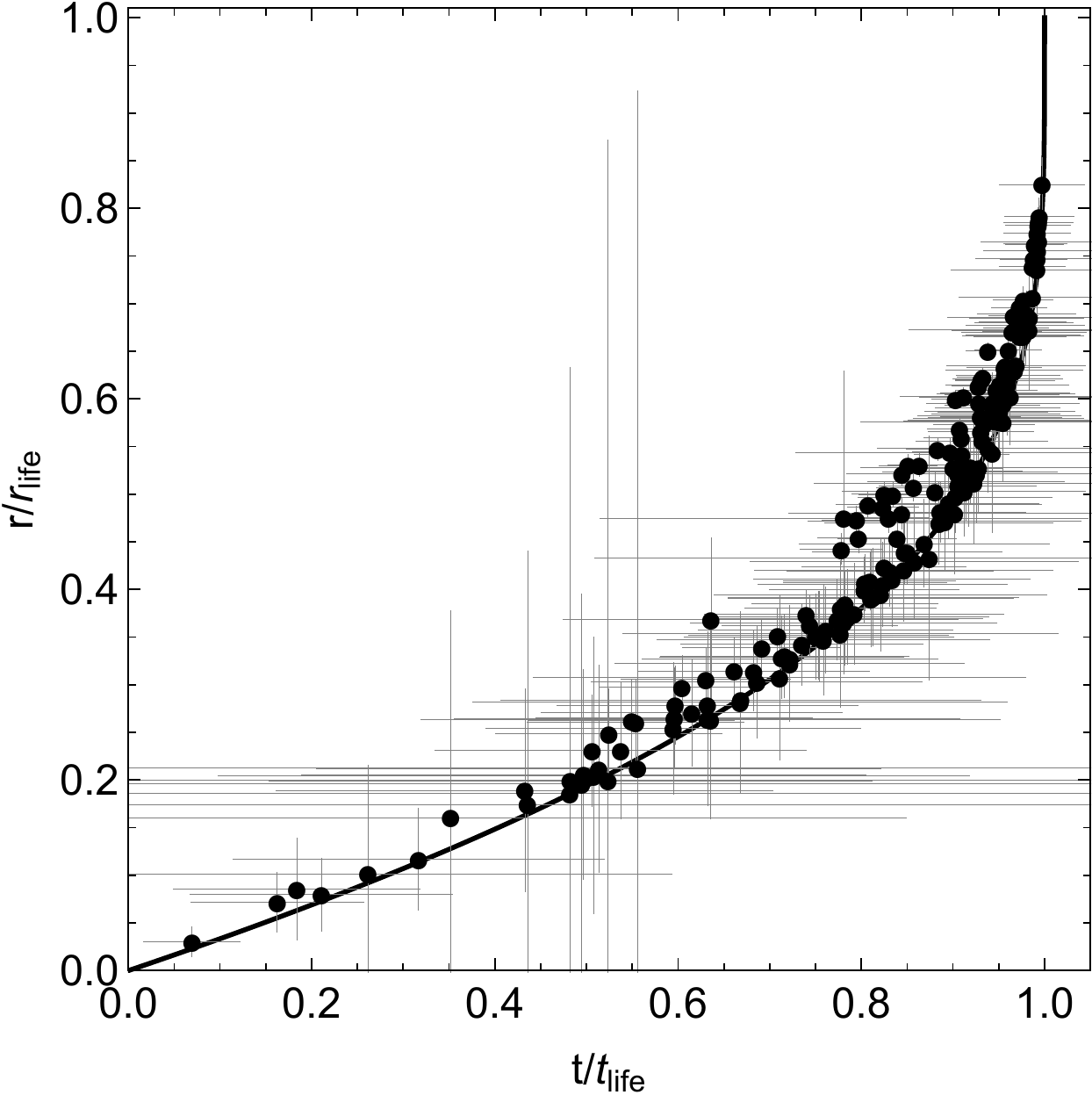}
\caption[Actual distance and age compared to the final distance and lifetime]{Distance normalized with the stopping distance versus time normalized with the lifetime for each data from Figure~\ref{fig:rVdatos}. The black line corresponds to the prediction using a $0.025M_\odot$ plasmon.} 
\label{fig:rftf}
\end{figure}

Finally, in Figure~\ref{fig:rftf} we show the time and position of each clump compared with its own lifetime and 
stopping distance, respectively. Again, there is a tendency for the most of the clumps to be at the end of their 
lives. This suggests that maybe some fingers have already ended their lives, explaining that there are H$_2$ features 
with no proper motion and CO streamers with no H$_2$ fingers associated. This characteristic can be explained in terms of extinction, but the radial velocities of the H$_2$ fingers are needed in order to correctly associate them to the CO streamers.

\section{Conclusions}
\label{sec:conclusions}

The plasmon model is a useful tool for the analysis of the dynamics of a clump interacting with a dense environment. 
Using the dynamic models presented in DA and RO19 we estimate the physical features, initial velocities and  masses, for the 
components (clumps, [Fe{\small II}] and HH object) reported in \cite{LB00}, \cite{DETAL02} and \cite{CPHD06}.

We obtain that the individual maximum mass for the clumps
is $0.2$~M$_\odot$, but the maximum velocity of this sample is of 800~km~s$^{-1}$. The total kinetic energy, in 
this case, is $\sim 3\times10^{49}$~erg, which represents $10^2$ times more 
energy than the energy obtained for the total luminosity in the Orion Fingers region. 

Other two consequences of the plasmon model is that the larger ejection velocities produce the shorter 
lifetimes, and the initial mass of a clump determines its stopping distance. The RO19 plasmon predicts that the longest fingers in Orion BN/KL have almost reached their lifetime, but they are not far from their final length and they required ejection velocities as high as 
800~km~s$^{-1}$ to reproduce the observations. This implies that the slower fingers could have lifetimes as long as 3000~yr, and the explosion signatures could disappear in 2000~yr.
The mass-loss plasmon can explain that there are not visible longer fingers because, if there were clumps thrown with higher speed or less mass, 
they could have died by now. Also, the required ejections velocities for most of the longest fingers are about 500~km~s$^{-1}$ which 
is less than twice their observed velocity.

Therefore, using the RO19 model we obtained the initial masses of each of the clumps,
from their mass distribution it is observed a large quantity of clumps has a mass in the interval of $8\times 10^{-3} - 2\times 10^{-1}$~M$_\odot$ and from the
velocities distribution, we obtain a distribution of 2 populations, one of them with a maximum at 200~km~s$^{-1}$ and another with a velocity of 500~km~s$^{-1}$.

Finally, from our calculated time and position of each clump and their own 
expected lifetime we can see a tendency for the most of the clumps to be at the end of their 
lives. We proposed that some fingers have already ended their lives, it explains that there are H$_2$ features  with no proper motion and CO streamers with no 
H$_2$ fingers associated.

%
\acknowledgments
We acknowledge support from PAPIIT-UNAM grants IN-109518 and IG-100218. P.R.R.-O. acknowledges
scholarship from CONACyT-M\'exico and financial support from COZCyT.
L.A.Z. acknowledge financial support from DGAPA, UNAM, and CONACyT, M\'exico. The authors thank Dr. Bally for his useful comments to improve this manuscript.
%







\end{document}